\documentclass[onecolumn,preprintnumbers,eqsecnum,superscriptaddress,10pt,floatfix,amsmath,amssymb]{revtex4}
\usepackage{float}
\usepackage{graphicx}
\usepackage{epsf}
\usepackage{dcolumn}   
\usepackage{bm}        
\usepackage{color}

\newcommand{\be}{\begin{equation}}
\newcommand{\en}{\end{equation}}
 \newcommand{\bea}{\begin{eqnarray}}
 \newcommand{\ena}{\end{eqnarray}}

\begin{document}

\title{P-V criticality in the extended phase space of black holes in Einstein-Horndeski gravity}

\author{Ya-Peng Hu}\email{huyp@nuaa.edu.cn}
\address{College of Science, Nanjing University of Aeronautics and Astronautics, Nanjing 210016, China}
\address{Center for Gravitation and Cosmology, College of Physical Science and Technology, Yangzhou University, Yangzhou 225009, China}

\author{Hong-An Zeng}\email{hazeng@nuaa.edu.cn}
\address{College of Science, Nanjing University of Aeronautics and Astronautics, Nanjing 210016, China}

\author{Zhi-Ming Jiang}\email{jzm931106lj@gmail.com}
\address{College of Science, Nanjing University of Aeronautics and Astronautics, Nanjing 210016, China}

\author{Hongsheng Zhang }\email{sps_zhanghs@ujn.edu.cn}
\address{School of Physics and Technology, University of Jinan, 336 West Road of Nan Xinzhuang, Jinan, Shandong 250022, China}

\begin{abstract}
  Gravity is believed to have deep and inherent relation to thermodynamics. We study phase transition and critical behavior in the extended phase space of asymptotic anti de-Sitter (AdS) black holes in Einstein-Horndeski gravity. We demonstrate that the black hole in Einstein-Horndeski gravity undergo phase transition and P-V criticality mimicking the van der Waals gas-liquid system. The key approach in our study is to introduce a more reasonable pressure instead of previous pressure $P=-\Lambda/8\pi$ related to cosmological constant $\Lambda$, and this proper pressure is given insight from the asymptotical behaviour of this black hole. Moreover, we also first obtain P-V criticality in the two cases with $\Lambda=0$ and $\Lambda>0$ in our paper, which implicates that the cosmological constant $\Lambda$ may be not a necessary pressure candidate for black holes at the microscopic level. We present critical exponents for these phase transition processes.

PACS number: 04.70.Dy, 04.20.-q, 04.50.Kd

\end{abstract}
\keywords{P-V criticality; Einstein-Horndeski gravity; phase transition; criticality exponents}

\maketitle



\section{Introduction}
The relation between thermodynamics and gravity theory is an interesting and profound issue. In the 1970s, the four thermodynamics laws of a stationary black hole system (in fact, spacetime thermodynamics, because the physical quantities in black hole thermodynamics should be treated as the quantities of the globally asymptotic flat manifold) was set up in \cite{Bardeen} and confirmed by Hawking radiation \cite{Hawking}. After this pioneer discovery of four thermodynamics laws in black hole system, more and more analogies are found between ordinary thermodynamics and black hole thermodynamics, including several phase transition and critical phenomena. A well known case is the so-called Hawking-Page phase transition~\cite{HP}, where a first order phase transition has been found to occur in the Schwarzschild-AdS black hole spactime. In addition, the phase transitions between radiation gas and a black hole have also been further investigated in an isolated box~\cite{Gibbons:1976pt,Hu:2017nzw}, where an equilibrium between the absorbed and radiated particles in a black hole system can be reached. Moreover, the phase transition similar to Van der Waals gas/liquid transition has been found in the Reissner-Nordstrom AdS (RN-AdS) black hole~\cite{Chamblin:1999hg}. These studies reveal significant properties of gravity systems. Interestingly and particularly, it may be helpful to solve some extremely difficult problems beyond gravity through investigations of phase transitions in gravity system. For example, the Hawking-Page phase transition maps to quark confinement/deconfinement transition in the scenario of AdS/CFT \cite{witten}, which has gained much more physical significance.

In recent years, a more similar phase transition mimicking the Van der Waals gas/liquid transition in the Reissner-Nordstrom AdS (RN-AdS) black hole system has been further investigated in \cite{Kubiznak:2012wp}. Because in fact no pressure P or volume V are defined in the previous work~\cite{Chamblin:1999hg}, while we use the P-V diagram to characterize the van der Waals liquid-gas phase transition. In~\cite{Kubiznak:2012wp}, a crucial step is to introduce a thermodynamic pressure P, which is proportional to the negative cosmological constant $\Lambda$ in the Reissner-Nordstrom AdS black hole system as $P=-\Lambda/8\pi$. Hence, this phase transition is also called P-V criticality of black holes, which has presented the more similarity between P-V diagrams of a black hole system and the Van der Waals-Maxwell gas/liquid system. Note that, this behavior of P-V criticality has been already found in more different black hole systems with asymptotical AdS behaviour~\cite{Gunasekaran:2012dq,Hendi:2012um,Zhao:2013oza,Wei:2012ui,Spallucci:2013osa,Miao:2018fke,Cai:2013qga,Xu:2015rfa,Majhi:2016txt}. Moreover, under this situation of pressure P proportional to the negative cosmological constant, the mass of a black hole becomes enthalpy rather than energy, which is required by thermodynamic laws~\cite{Kastor:2009wy}. In these laws, the term $dP$ appears, which implies that the cosmological 'constant' may be a variable, and there have been a large amount of discussions of a variable vacuum energy in contexts of cosmology since the discovery of cosmic acceleration. Theoretically, we have several motivations to invoke the assumption of  a variable  cosmological 'constant'. If the present theory is not a 'fundamental theory', the constants in the theory may be variables. For example, the cosmological constant becomes a variable in gauged supergravity. And furthermore, the cosmological constant, Newton constant, and other coupling constants may become variables in quantum theory, whose vacuum expectations are the observed values~\cite{vacexp}. Thus it may be reasonable to identify the cosmological constant as the pressure in P-V criticality in gravity and allow the pressure to be a variable. Such an identity leads to several sensible results~\cite{Kubiznak:2012wp,Gunasekaran:2012dq,Hendi:2012um,Zhao:2013oza,Wei:2012ui,Spallucci:2013osa,Miao:2018fke,Cai:2013qga,Xu:2015rfa,Majhi:2016txt}. However, it still needs more examinations, since the $P-V$ criticality of gravity system compared to gas/liquid system is fundamentally an analogy. We have no sound statistical mechanics behind the gravity thermodynamics, and thus we are unable to prove that the pressure $P$ should be identified as the cosmological constant $\Lambda$ at the microscopic level such as $P=-\Lambda/8\pi$.

The scalar-tensor gravity has a fairly long history and several different extensions, while the Horndeski scalar tensor theory is the most generic scalar tensor theory. In Horndeski scalar tensor theory, the equations of motion consist of terms which have at most second-order derivatives acting on each fields, though it permits higher-order derivatives in the action \cite{horn}. This property is very similar to the Lovelock gravity. From different considerations, the Horndeski theory is rediscovered in a different form in the studies of cosmology\cite{Deffayet:2011gz,Gao:2014soa}. Black holes solutions in a special Einstein-Horndeski gravity are derived in \cite{bhhorn,Cisterna:2014nua}, while thermodynamics for black holes in Einstein-Horndeski is investigated in \cite{Feng:2015wvb,Miao:2016aol}. In our paper, we shall show that there is no P-V criticality in black holes in Einstein-Horndeski theory, if one just treat the cosmological constant as pressure as usual. We demonstrate that the proper pressure is different from the cosmological constant. With this new pressure, the P-V criticality appears. Moreover, we also first find out P-V criticality in the two cases with $\Lambda=0$ and $\Lambda>0$, which implicates that the cosmological constant $\Lambda$ may be not a necessary pressure candidate for black holes at the microscopic level.

This article is organized as follows. In the next section, we briefly review an exact solution in the Einstein-Horndeski theory. In section III, we demonstrate that P-V phase transition does occur with a new definition of the pressure different from the cosmological constant. In section IV, we conclude this paper.

\section{Thermodynamics of static black hole solutions in Horndeski gravity}
The most general action of Horndeski gravity can be seen in the reference~\cite{Gao:2014soa}. In our paper, we just investigate a special case in Horndeski gravity with a non-minimal kinetic coupling, and the corresponding action is written as~\cite{Cisterna:2014nua,Feng:2015wvb,Miao:2016aol}
\begin{eqnarray}
S=\int\sqrt{-g}  d^4x  [(R-2\Lambda)- \frac{1}{2} (\zeta g_{\mu \nu}-\eta G_{\mu \nu}) \nabla^{\mu} \phi \nabla^{\nu} \phi - \frac{1}{4} F_{\mu \nu} F^{\mu \nu}], \label{action}
\end{eqnarray}
where $\zeta$ is a coupling constant, and $\eta$ stands for the coupling strength with the dimension of length square. Besides the electromagnetic field $F_{\mu\nu}$, a real scalar field $\phi$ also exists, which has a non-minimal kinetic coupling with the metric and Einstein tensor. From this action, the equations of motion are
\begin{eqnarray}
G_{\mu\nu}+\Lambda g_{\mu\nu}&=&\frac12 \left(\zeta T_{\mu\nu}+\eta \Xi_{\mu\nu}+E_{\mu\nu}\right), \label{gmu}\\
\nabla_{\mu}\left[(\zeta g^{\mu\nu}-\eta G^{\mu\nu})\nabla_{\nu}\phi\right]&=&0, \label{scalarmu}\\
\nabla_{\mu}F^{\mu\nu}&=&0, \label{emu}
\end{eqnarray}
where $T_{\mu\nu}$, $\Xi_{\mu\nu}$, and $E_{\mu\nu}$ are defined as
\begin{eqnarray}
T_{\mu\nu} &\equiv& \nabla_{\mu}\phi \nabla_{\nu}\phi-\frac12 g_{\mu\nu}\nabla_{\rho}\phi \nabla^{\rho}\phi,\\
\Xi_{\mu\nu} &\equiv & \frac12 \nabla_{\mu}\phi \nabla_{\nu}\phi R-2\nabla_{\rho}\phi \nabla_{(\mu}\phi R_{\nu)}^{\rho}-\nabla^{\rho}\phi \nabla^{\lambda}\phi R_{\mu\rho\nu\lambda}\nonumber \\
& &-(\nabla_{\mu}\nabla^{\rho}\phi)(\nabla_{\nu}\nabla_{\rho}\phi)
+(\nabla_{\mu}\nabla_{\nu}\phi)\square \phi+\frac12 G_{\mu\nu}(\nabla \phi)^2\nonumber\\
& &-g_{\mu\nu}\left[-\frac12 (\nabla^{\rho}\nabla^{\lambda}\phi)(\nabla_{\rho}\nabla_{\lambda}\phi)
+\frac12 (\square \phi)^2-\nabla_{\rho}\phi \nabla_{\lambda}\phi R^{\rho\lambda}\right],\\
E_{\mu\nu}&\equiv& F_{\mu}^{\rho}F_{\nu\rho}-\frac{1}{4} g_{\mu\nu}F^2.
\end{eqnarray}
In our paper, we focus on the static black hole solutions with spherical symmetry to Eqs.~(\ref{gmu})-(\ref{emu}), while the metric, scalar field and Maxwell field are simplified
\begin{eqnarray}
\text{d} s^2&=&-f(r)\text{d} t^2+g(r)\text{d} r^2+r^2(\text{d} \theta^2+\sin^2 \theta \text{d}\varphi^2), \label{metric}\\
\phi&=&\phi(r),~~ A=\Psi(r) \text{d}t,
\end{eqnarray}
Through this simplification, an analytical solution has been obtained~\cite{Cisterna:2014nua,Feng:2015wvb,Miao:2016aol}
\begin{eqnarray}
f(r)&=&\frac{\zeta r^2}{3\eta}-\frac{2M}{r}+\frac{3\zeta+\Lambda \eta}{\zeta-\Lambda \eta}+\left(\frac{\zeta+\Lambda \eta+\frac{\zeta^2 q^2}{4\eta}}{\zeta-\Lambda \eta}\right)^2\,\frac{\tan^{-1}\left(\sqrt{\frac{\zeta}{\eta}}r\right)}{\sqrt{\frac{\zeta}{\eta}}r}\nonumber \\
& &+\frac{\zeta^2 q^2}{(\zeta-\Lambda \eta)^2 r^2}-\frac{\zeta^2 q^4}{48(\zeta-\Lambda \eta)^2 r^4}+\frac{\zeta^3 q^4}{16\eta(\zeta-\Lambda \eta)^2 r^2},\nonumber\\
g(r)&=&\frac{\zeta^2[4(\zeta-\Lambda \eta) r^4+8\eta r^2-\eta q^2]^2}{16r^4(\zeta-\Lambda \eta)^2(\zeta r^2+\eta)^2 f(r)},\nonumber\\
\psi^2(r)&=&-\frac{\zeta^2[4(\zeta+\Lambda \eta)r^4+\eta q^2][4(\zeta-\Lambda \eta) r^4+8\eta r^2-\eta q^2]^2}{32\eta r^6(\zeta-\Lambda \eta)^2 (\zeta r^2+\eta)^3 f(r)},\nonumber\\
\Psi(r)&=&\Psi_0+\frac14 \frac{q\sqrt{\zeta}}{\eta^{\frac32}}\left[\frac{4\eta(\zeta+\Lambda\eta)+\zeta^2 q^2}{\zeta-\Lambda\eta}\right]\tan^{-1}\left(\frac{1}{\sqrt{\frac{\zeta}{\eta}}r}\right)\nonumber \\
& &-\frac{\zeta q(\zeta q^2+8\eta)}{4\eta r(\zeta-\Lambda\eta)}+\frac{\zeta q^3}{12r^3(\zeta-\Lambda\eta)}, \label{elepo}
\end{eqnarray}
where $\psi(r) \equiv \phi^{\prime}(r)$ and $\Psi_0$ is an integration constant. Note that, this solution (\ref{elepo}) is same as that in Ref.~\cite{Feng:2015wvb} after taking identity $\tan^{-1} x=\frac{\pi}{2}-\tan^{-1}(1/x)$ into account, and both $\zeta$ and $\eta$ can not be zero during obtaining this solution \protect\footnotemark. Therefore, we can not switch off the scalar field within this solution family, which implies that this solution is usually not continuously connected with the maximally symmetric background, and hence does not contain the Reissner-Nordstom-AdS solution. Despite of this, the case $M=0$ and $q=0$ is also a regular spacetime, which can describe an asymptotically AdS gravitational soliton solution~\cite{Cisterna:2014nua}. Particularly, if we further require $\zeta=-\Lambda \eta$, this asymptotically AdS gravitational soliton solution becomes the pure AdS solution. Hence, in this particular case, we can find that the Schwarzschild-AdS solution is recovered with $\zeta=-\Lambda \eta$ and $q=0$. However, the Reissner-Nordstom-AdS solution still can not be recovered in this particular case. In our paper, we just consider this solution (\ref{elepo}) with an asymptotical AdS behavior, and hence we further require $\zeta /\eta >0$ and $\zeta \neq \Lambda \eta$.  The temperature of this black hole in Horndeski gravity is easily obtained
\footnotetext{For the equation of motion for the scalar field, one can obtain Eq.(10) in Ref.~\cite{Cisterna:2014nua} or Eq.(3.1) in Ref.~\cite{Feng:2015wvb}. From this equation, obviously $\eta$ can not be zero. For the case $\zeta \neq 0$, one will obtain the solution (\ref{elepo}). For the particular case $\zeta=0$, Ref.~\cite{Cisterna:2014nua} also discussed it in Sec V. However, this case can not be contained in this solution (\ref{elepo}).}
\begin{align}
T=\frac{1}{4\pi }\sqrt{g'\left(r_h\right)f'\left(r_h\right)}=\frac{1}{4\pi  r_h}\left(\frac{2\zeta }{\zeta -\Lambda \eta }+\frac{\zeta r_h^2}{\eta}-\frac{\text{$\zeta $q}^2 }{4r_h^2(\zeta -\Lambda \eta)}\right), \label{Temperature}
\end{align}
where $r_h$ is the location of outer horizon satisfying $f\left(r_h\right)=0$, and the first law of thermodynamics of this black hole solution (\ref{elepo}) has been carefully investigated in Ref.\cite{Feng:2015wvb}
\begin{eqnarray}
dE=TdS+\Phi_e dQ_e+\Phi_\chi dQ_\chi, \label{FirstLaw}
\end{eqnarray}
where energy $E$, entropy $S$, charge potential $\Phi_e$ , charge $Q_e$, scalar charge $Q_\chi$ and its conjugate potential $\Phi_\chi$ are calculated as
\begin{eqnarray}
E &=&\frac{M(\zeta -\eta  \Lambda) }{2 \zeta } -\frac{\pi  [q^2 \zeta ^2+4 \eta  (\zeta +\eta  \Lambda )]^2}{128
\zeta ^{3/2} \eta ^{3/2} (\zeta -\eta  \Lambda )},~~S=\frac{2\pi ^2 r_h^3 (\zeta - \eta  \Lambda )}{\zeta \left(1+r_h^2 \frac{\zeta }{\eta }\right)}T,~~\Phi_e=\Psi_0, \nonumber\\
Q_e &=&\frac{q}{4},~~Q_{\chi} =2 \sqrt{2} \pi  \sqrt{-\frac{q^2+4 r_h^4 \left(\frac{\zeta }{\eta }+ \Lambda \right)}{r_h^2 \left(\eta +r_h^2
\zeta \right)}},~~\Phi_{\chi} =-\frac{ r_h^2 T \eta }{32 \pi }Q_{\chi}. \label{Thermo-quanties}
\end{eqnarray}
Interestingly, due to the scalar charge $Q_{\chi}$ on the horizon, the entropy in (\ref{Thermo-quanties}) disagrees with the Bekenstein-Hawking entropy, where the entropy is explicitly calculated from Wald formalism as $S=\left(1+\frac{\eta}{4}(\frac{\phi'(r)^{2}}{g(r)})|_{r_h}\right)\frac{A}{4}$ in Ref.\cite{Feng:2015wvb}, and $A=4\pi r_h^2$ is the horizon area of black hole.

Note that, during investigations of P-V criticality of asymptotical AdS black hole system, the cosmological constant $\Lambda$ is usually considered as a dynamical variable. In our case, parameters $\zeta$ and $\eta$ in the Lagrangian of action (\ref{action}) can be also considered as dynamical variables. However, due to absence of scalar field potential in the action, in fact we just have one independent parameter for these two parameters $\zeta$ and $\eta$, i.e. parameter $\zeta$ or $\eta$ can be set to unity by the redefinition of $\phi$. Therefore, for simplicity and without loss of generality, we just consider $\zeta$ as the one independent parameter and fix $\eta$ as a constant in the following\protect\footnotemark. In this case, the first law of thermodynamics of black hole solution in (\ref{FirstLaw}) is further investigated as
\begin{align}
dE=TdS+\Phi_e dQ_e+\Phi_\chi dQ_\chi+V_A d \Lambda +V_B d P_B, \label{NewFirstLaw}
\end{align}
 where $V_A$ is the corresponding conjugate to $\Lambda$, and $V_B$ is the corresponding conjugate to $P_B\equiv\zeta/(8\pi\eta)$, which have been explicitly shown in the appendix~\ref{A}.
\footnotetext{Note that, for the case $\eta$ as the one independent parameter with constant $\zeta$, volume $V_B$ will be changed. However, changes of $V_B$ do not affect the detailed pressure function $P(r_h, T)$ (\ref{p2}) in the following. Hence, two cases have the same investigations of P-V criticality.}

In addition, from the formula of temperature in (\ref{Temperature}), it can be further rewritten as
\begin{align}
4r_h{}^4\Pi ^{*2}+\left(8r_h{}^2-4\text{$\Lambda $r}_h{}^4-16\pi  T r_h{}^3-q^2\right)\Pi ^*+16\pi  T r_h{}^3\Lambda =0, \label{pvt}
\end{align}
where a new parameter $\Pi ^*\equiv\frac{\zeta }{\eta }$  has been defined for convenience of later investigations.

\section{P-V criticality in the extended phase space of black holes in Horndeski gravity }
There have been a lot of works to investigate P-V criticality in the extended phase spaces of asymptotical AdS black holes~\cite{Kubiznak:2012wp,Gunasekaran:2012dq,Hendi:2012um,Zhao:2013oza,Wei:2012ui,Spallucci:2013osa,Miao:2018fke,Cai:2013qga,Xu:2015rfa,Majhi:2016txt}. Since the black hole solution in (\ref{elepo}) also has asymptotical AdS behaviour~\cite{Cisterna:2014nua,Feng:2015wvb,Miao:2016aol}, we will investigate the P-V criticality in its extended phase space in Horndeski gravity.

\subsection{No P-V criticality with Pressure defined as $P=-\frac{1}{8\pi }\Lambda$} \label{NorPressure}
Among many previous investigations on $P$-$V$ criticality in extended phase spaces of black holes with asymptotical AdS behaviour ~\cite{Kubiznak:2012wp,Gunasekaran:2012dq,Hendi:2012um,Zhao:2013oza,Wei:2012ui,Spallucci:2013osa,Miao:2018fke,Cai:2013qga,Xu:2015rfa,Majhi:2016txt}, the thermodynamic pressure $P$ of a black hole system is usually defined as
\begin{equation}
P=-\frac{1}{8\pi }\Lambda \label{P}.
\end{equation}
Therefore, as the corresponding conjugate to $\Lambda$, $V_A$ is easily seen to be proportional to thermodynamic volume of this black hole system from (\ref{NewFirstLaw}). In this subsection, we will check the possibility of $P$-$V_A$ criticality by using this definition of pressure.

\begin{figure}[H]
\centering
\includegraphics[width=.6\textwidth]{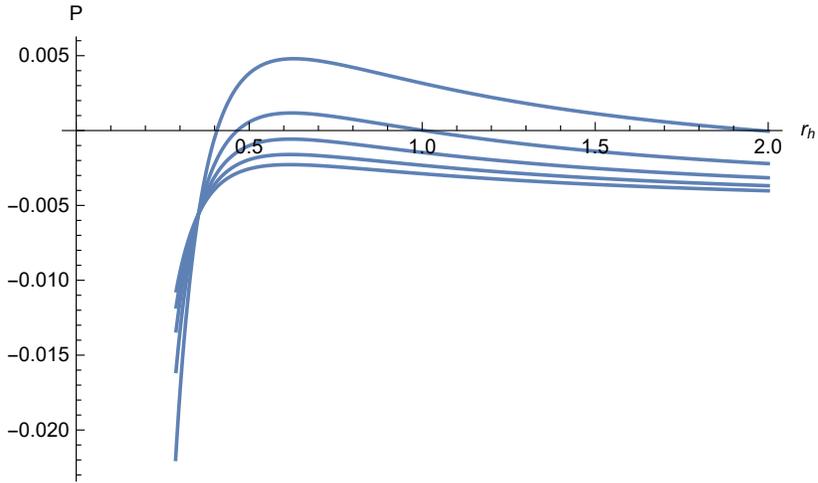}
\caption{$q=1, \Pi^*=0.14$, from top to bottom, $T=0.15$ to $0.35$ with interval 0.05\protect\footnotemark} \label{fig5}
\end{figure}

Note that, the $P$-$V$ criticality is also usually discussed from investigations on the pressure function $P(r_h, T)$ through $P-r_h$ diagram in previous works~\cite{Kubiznak:2012wp,Gunasekaran:2012dq,Hendi:2012um,Zhao:2013oza,Wei:2012ui,Spallucci:2013osa,Miao:2018fke,Cai:2013qga,Xu:2015rfa,Majhi:2016txt}. Since $V_A$ is complicated in our case, we also investigate the $P-V$ criticality through discussions on the corresponding function $P(r_h, T)$ in the $P-r_h$ phase diagram. In appendix~\ref{B}, we have also given a simple proof to show the equivalence between investigations on $P-V$ criticality through $P-V$ and $P-r_h$ phase diagrams. For the pressure function $P(r_h, T)$, after considering (\ref{P}) in (\ref{pvt}), we easily obtain
\begin{align}
P(r_h,T)=\frac{4r_h{}^4\Pi ^{*2}+\left(8r_h{}^2-16\pi  T r_h{}^3-q^2\right)\Pi ^*}{8\pi \left(16\pi  T r_h{}^3-4r_h{}^4\Pi ^*\right)}, \label{Pressure2}
\end{align}
which is complicated to analytically investigate whether there is $P$-$V$ criticality like the van der Waals liquid-gas system. Therefore, we check the criticality through plotting the $P-r_h$ phase diagram for different cases, and we find that no evidence shows the existence of $P$-$V$ criticality in this case, since the $P-r_h$ phase diagrams are all similar to the following diagram in Fig.~\ref{fig5}
\footnotetext{Strictly speaking, regions with positive pressure in the diagram are physical here.}

\subsection{P-V criticality with a more reasonable Pressure defined as $P=\frac{\zeta}{8 \pi \eta}$} \label{NPressure}
In the above subsection, we have found that there is no P-V criticality with pressure defined as $P=-\frac{1}{8\pi }\Lambda$. However, for the solution in (\ref{elepo}), it is an asymptotical AdS solution with an effective AdS radius $l_{eff}$ satisfying $l_{eff}^2=\frac{3\eta}{\zeta}$, which is independent on the cosmological constant $\Lambda$. Therefore, a more reasonable pressure is defined as
\begin{equation}
P=-\frac{\Lambda_{eff}}{8\pi}=\frac{\zeta}{8\pi \eta}=\frac{\Pi^*}{8\pi}=P_B, \label{NewP}
\end{equation}
where $\Lambda_{eff}$ is an effective cosmological constant related to $l_{eff}$ as $\Lambda_{eff}=-\frac{3}{l_{eff}^2}=-\frac{\zeta}{\eta}$. In this subsection, we will check the possibility of $P-V$ criticality by using this new reasonable pressure in (\ref{NewP}). Note that, pressure $P$ is a dynamical variable in the $P-V$ criticality, while there are two parameters $\zeta$ and $\eta$ in the new defined pressure (\ref{NewP}). For the simplicity, we just consider the parameter $\zeta$ as a dynamical variable in the following, while the parameter $\eta$ is fixed as a constant. In this case, similar to the above subsection, $V_B$ is also easily seen to be a new thermodynamic volume of this black hole system from (\ref{NewFirstLaw}), and here this new thermodynamic volume is the conjugate to this new defined pressure. In the following, we will also use the pressure function $P(r_h, T)$ and plot $P-r_h$ phase diagrams to investigate the $P-V$ criticality. A simple proof of equivalence between investigations on $P-V$ criticality through $P-V_B$ or $P-r_h$ phase diagrams is presented in appendix~\ref{B}, too.

As a new pressure function $P(r_h,T)$ corresponding to the new defined pressure (\ref{NewP}), it is also solved from (\ref{pvt}), and one obtains two branches
\begin{align}
P(r_h,T)=\frac{-(8r_h{}^2-4\text{$\Lambda $r}_h{}^4-16\pi  T r_h{}^3-q^2)\pm\sqrt{\left(8r_h{}^2-4\text{$\Lambda $r}_h^4-16\pi
 T r_h{}^3-q^2\right){}^2-256\pi \Lambda r_h^7 T }}{64\pi r_h^4}.\label{p2}
\end{align}

For the case with zero cosmological constant $\Lambda=0$, one branch becomes $P(r_h,T)=0$, while the other is
\begin{eqnarray}
P(r_h,T)=-\frac{8r_h{}^2-16\pi  T r_h{}^3-q^2}{32\pi r_h^4}=\frac{\omega _1}{r_h}+\frac{\omega _2}{r_h{}^2}+\frac{\omega _4}{r_h{}^4}, \label{LikeMGcase}
\end{eqnarray}
where
\begin{eqnarray}
\omega _1=\frac{T}{2}, \quad \omega _2=-\frac{1}{4\pi }, \quad \omega _4=\frac{q^2}{32\pi }.
\end{eqnarray}
Note that this nonzero pressure function $P(r_h,T)$ (\ref{LikeMGcase}) is smilar to the ones in RN-AdS black hole~\cite{Kubiznak:2012wp} and massive gravity~\cite{Xu:2015rfa}. Moreover, we can also identify specific volume $v$ of this black hole system with horizon radius as $v=2r_h$~\cite{Kubiznak:2012wp}. Therefore, an analytical investigation on the P-V criticality of this black hole system is similar to those in ~\cite{Kubiznak:2012wp} and ~\cite{Xu:2015rfa}, and hence the critical point is obtained~\cite{Xu:2015rfa}
\begin{eqnarray}
r_{hc}=\sqrt{-\frac{6w_4}{w_2}},~w_{1c}=-\frac{4}{3}w_2\sqrt{-\frac{w_2}{6w_4}},~P_c=\frac{w_2^2}{12w_4},
\end{eqnarray}
while the corresponding critical exponents around the critical point are $\alpha =0,\beta =\frac{1}{2},\gamma =1,\delta =3$. These critical exponents are same as those in~\cite{Kubiznak:2012wp,Xu:2015rfa}.

\begin{figure}[H]
\centering
\includegraphics[width=.6\textwidth]{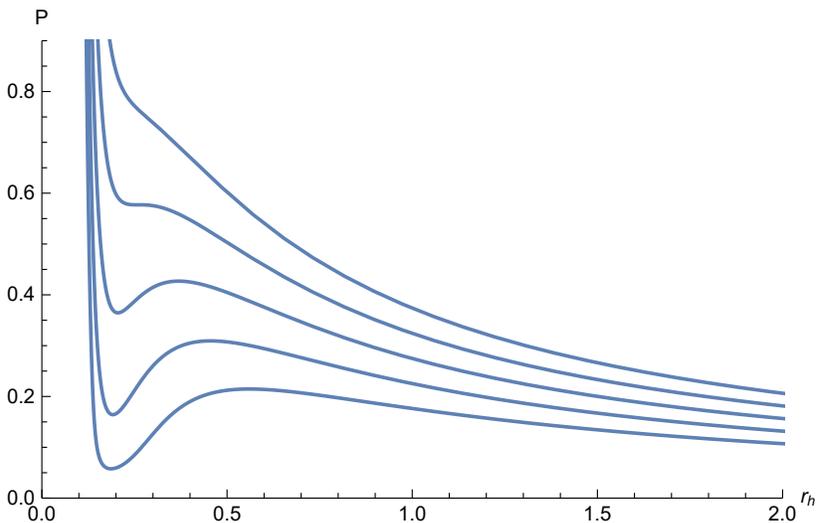}
\caption{$q=0.3,\Lambda=-0.3$, from bottom to top, $T=0.6$ to $1.0$ with interval 0.1 }\label{fig33}
\end{figure}

\begin{figure}[H]
\centering
\includegraphics[width=.6\textwidth]{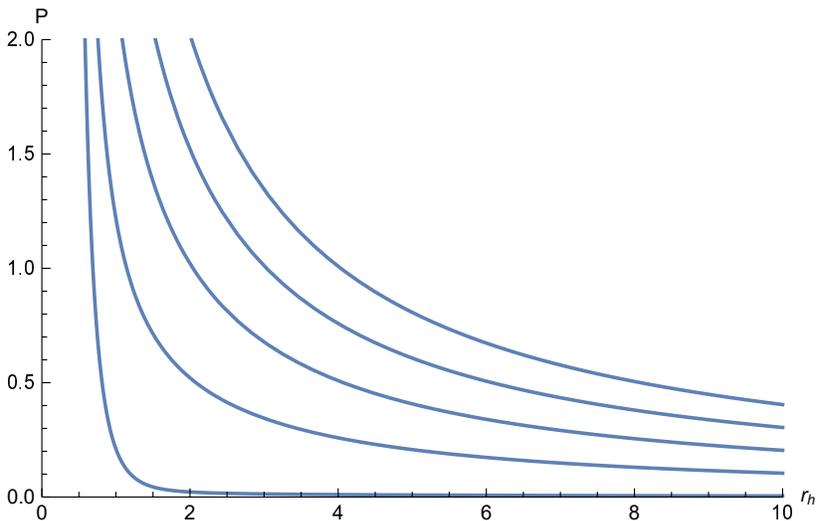}
\caption{$q=5,\Lambda=-0.3$, from bottom to top, $T=1.1$ to $9.1$ with interval 2.0}\label{fig44}
\end{figure}

For the case with negative cosmological constant $\Lambda < 0 $, we should choose the $'+'$ branch of function $P(r_h,T)$ in (\ref{p2}), since pressure in $'-'$ branch is negative. In this $'+'$ branch, the analytical investigation of $P$-$V$ criticality is complicated, because the pressure function $P(r_h,T)$ in (\ref{p2}) is complicated. For simplicity, we just plot function $P(r_h,T)$ to find out its P-V criticality in this '+' branch. Indeed, we have found out that there is P-V criticality for some fixed parameters, e.g. parameters with $q=0.3,\Lambda=-0.3$ in Fig.~\ref{fig33}
However, it should be pointed out that the P-V criticality does not occur for all cases with fixed parameters. For example, in Fig.~\ref{fig44}, there is no evidence to show that the case with parameters $q=5,\Lambda=-0.3$ has P-V criticality

For the P-V criticality in the case with fixed parameter $q=0.3,\Lambda=-0.3$, we turn to calculate corresponding critical exponents in the following. First, the critical point is determined as the inflection point in the $P-r_h$ diagram, i.e.,
\begin{equation}\label{cricondition}
\frac{\partial P}{\partial
r_h}\Big|_{r_h=r_{hc},T=T_c}=\frac{\partial^2P}{\partial
r_h^{~2}}\Big|_{r_h=r_{hc},T=T_c}=0.
\end{equation}
From these two equations, we numerically obtain the critical point as $r_{hc}\approx 0.2598, T_c \approx 0.8002$ and $P_c \approx 0.5775$.

On the other hand, as same as the usual thermodynamic system, critical exponents $\alpha$, $\beta$, $\gamma$, and $\delta$ are defined as follows
\begin{eqnarray}
&&C_v \sim |t|^{-\alpha}~,
\nonumber
\\
&& \Delta_v \sim v_l-v_s\sim |t|^{\beta}~,
\nonumber
\\
&& K_T = -\frac{1}{v}\Big(\frac{\partial v}{\partial P}\Big)_T \sim |t|^{-\gamma}~,
\nonumber
\\
&&P-P_c \sim |v-v_c|^{\delta}~,
\label{exponent}
\end{eqnarray}
where $t=T/T_c-1$, $C_v$ is isopyknic heat capacity, $v$ stands for specific
volume of black hole with $v=2r_h$, and $\kappa_T$ is isothermal compressibility. The subscript
$c$ stands for critical point, while $l$ and $s$ refer to large and small black hole phases similar to the gas and liquid phases of the usual thermodynamic system.

For the critical exponent $\alpha$, we need calculate the isopyknic heat capacity $C_v$ at the critical point. The entropy of this black hole (\ref{elepo}) has been obtained in~\cite{Feng:2015wvb} as
\begin{equation}
S=\frac{\pi\left(8r_h^2-4r_h^4(\Lambda-\Pi^*)-q^2\right)}{8(r_h^2\Pi^*+1)},   \label{Entropy}
\end{equation}
and hence the isopyknic heat capacity $C_v=T(\frac{\partial S}{\partial T})_v$ at the critical point is a constant. Therefore, we have $\alpha=0$.

For the critical exponent $\beta$, we first make the Taylor series expansion near critical point for the pressure function $P(r_h,T)$  in (\ref{p2})
\begin{equation}
P=P_c+Rt+Bt\omega+D\omega^3+Kt^2+..., \label{Pexpansion}
\end{equation}
where
\begin{equation}
t=\frac{T}{T_c}-1, \quad \omega=\frac{r_h}{r_{hc}}-1,
\end{equation}
and the expansion coefficients are numerically obtained as
\begin{align}
&R=T_c\left[\left(\frac{\partial P}{\partial T}\right)_{r_h}\right]_c \approx 1.4898, \quad
&B=T_c r_c \left[\left(\frac{\partial ^2P}{\partial T\partial r_h}\right)\right]{}_c \approx -1.4120, \nonumber\\
&D=r_c^3\frac{1}{3!}\left[\left(\frac{\partial ^3P}{\partial r_h{}^3}\right){}_T\right]{}_c \approx -0.7449, \quad
&K=T_c^2\frac{1}{2!}\left[\left(\frac{\partial ^2P}{\partial T^2}\right){}_{r_h}\right]{}_c \approx 0.1230.
\end{align}
For constant $t$, one finds $dP = (Bt+3D\omega^2)d\omega$, therefore, after using Maxwell's area law\cite{Kubiznak:2012wp,Xu:2015rfa,Majhi:2016txt}, we obtain the following equation
\begin{equation}
\int_{\omega_l}^{\omega_s}\omega(Bt+3D\omega^2)d\omega + \int_{\omega_l}^{\omega_s}(Bt+3D\omega^2)d\omega =0~,
\label{omega}
\end{equation}
where $\omega_l$ and $\omega_s$ correspond to the large and small volumes of black hole in two different phases. We further consider the condition $P_l=P_s$, i.e. the end point of vapor and the staring point of liquid have a same pressure like the usual thermodynamic liquid-gas system, and hence from (\ref{Pexpansion}) this condition implies
\begin{equation}
Bt(\omega_l-\omega_s) + D(\omega_l^3-\omega_s^3)=0~,
\label{first}
\end{equation}
which means that the second integral of (\ref{omega}) vanishes. Therefore, (\ref{omega}) is further reduced as
\begin{equation}
Bt(\omega_l^2-\omega_s^2)+\frac{3D}{2}(\omega_l^4-\omega_s^4) = 0~.
\label{second}
\end{equation}
For these two equations (\ref{omega}) and (\ref{second}), the non-trivial solutions are $\omega_l =(-Bt/D)^{1/2}$ and $\omega_s=-(-Bt/D)^{1/2}$. It is easy to find that
\begin{equation}
\Delta_v \sim v_l-v_s = (\omega_l-\omega_s)v_c\sim |t|^{1/2}~,
\end{equation}
which determines the critical exponent $\beta=1/2$.

For the critical exponent $\gamma$, the isothermal compressibility $K_T$ can be calculated by $(\partial P/\partial r_h)_T$ from (\ref{Pexpansion}), and it is given by
\begin{equation}
\Big(\frac{\partial P}{\partial r_h}\Big)_T \simeq \frac{B}{r_{hc}}t~,
\end{equation}
where $\partial \omega/\partial r_h= 1/r_{hc}$ has been used. Therefore, near the critical point, the value of $K_T$ is
\begin{equation}
K_T \simeq \frac{1}{Bt}\sim t^{-1}~,
\end{equation}
which implies $\gamma =1$. Finally, for the critical exponent $\delta$, when $T=T_c$, from (\ref{Pexpansion}), it yields
\begin{equation}
P-P_c\sim \omega^3\sim (r_h-r_{hc})^3~,
\end{equation}
which implies the value of the critical exponent $\delta$ is $\delta=3$.

\begin{figure}[H]
\centering
\includegraphics[width=.6\textwidth]{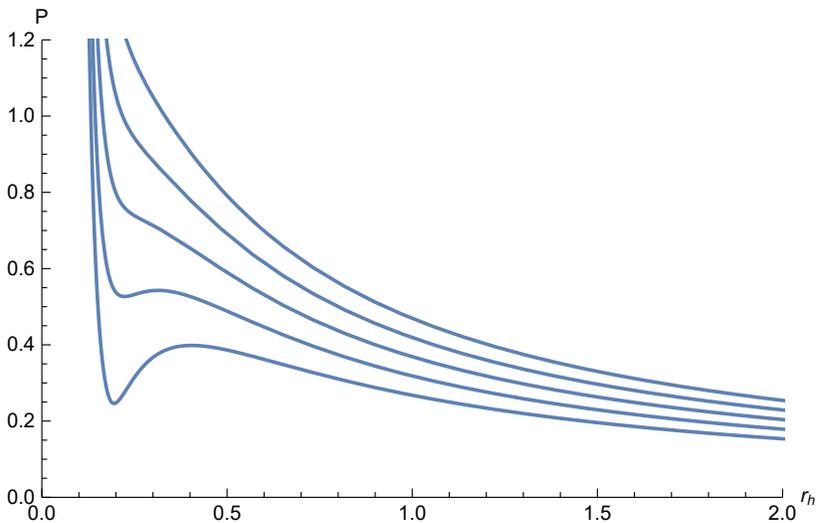}
\caption{$q=0.3, \Lambda=0.3$, from bottom to top, $T=0.8$ to $1.2$ with interval 0.1 } \label{fig3}
\end{figure}

\begin{figure}[H]
\centering
\includegraphics[width=.6\textwidth]{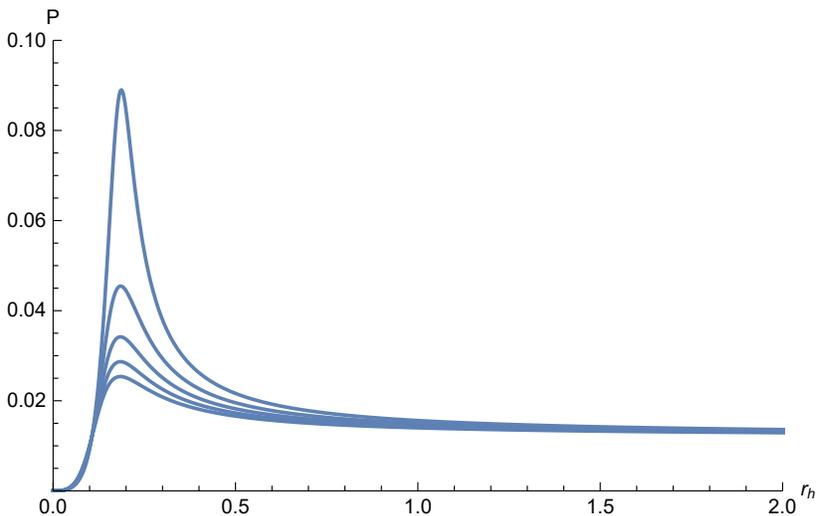}
\caption{$q=0.3, \Lambda=0.3$, from top to bottom , $T=0.8$ to $1.2$ with interval 0.1} \label{fig4}
\end{figure}

For the case with positive cosmological constant $\Lambda > 0 $, the two branches of pressure function $P(r_h,T)$ in (\ref{p2}) can be both chosen, while this constraint $8r_h{}^2-4\text{$\Lambda $r}_h{}^4-16\pi  T r_h{}^3-q^2<0$ should be satisfied to keep pressure $P(r_h,T)$ positive. For the $'+'$ branch, we find out that there is P-V criticality, and have plotted the phase diagram with $q=0.3, \Lambda=0.3$ in Fig.~\ref{fig3}
For the $'-'$ branch, there is no evidence to show its existence of P-V criticality. In Fig.~\ref{fig4}, we have plotted some phase diagrams as an illustration
Therefore, we just focus on the $'+'$ branch in the following, and calculate its corresponding critical exponents. For fixed parameters $q=0.3,\Lambda=0.3$, the critical point has been numerically obtained as $r_{hc}\approx 0.2598, T_c \approx 0.8333$ and $P_c \approx 0.6014$. Like the above procedure, we make the Taylor series expansion near the critical point for the pressure function in (\ref{p2}), while the corresponding expansion coefficients are
\begin{align}
&R=T_c\left[\left(\frac{\partial P}{\partial T}\right)_{r_h}\right]_c \approx 1.6598,
&B=T_c r_c \left[\left(\frac{\partial ^2P}{\partial T\partial r_h}\right)\right]{}_c \approx -1.7525,\nonumber\\
&D=r_c^3\frac{1}{3!}\left[\left(\frac{\partial ^3P}{\partial r_h{}^3}\right){}_T\right]{}_c \approx -0.8299,
&K=T_c^2\frac{1}{2!}\left[\left(\frac{\partial ^2P}{\partial T^2}\right){}_{r_h}\right]{}_c \approx -0.1632.
\end{align}
These four coefficients are also nonzero, and it is obvious that this case is similar to the above case with $q=0.3, \Lambda=-0.3$. Therefore, the critical exponents are same as the above case, i.e., $\alpha =0,\beta =\frac{1}{2},\gamma =1,\delta =3$.

Interestingly and obviously, all these derived critical exponents satisfy the following thermodynamic scaling laws
\begin{eqnarray}
&~&\alpha+2\beta+\gamma=2,~~~\alpha+\beta(1+\delta)=2,\nonumber\\
&~&\gamma(1+\delta)=(2-\alpha)(\delta-1),~~~\gamma=\beta(\delta-1),
\end{eqnarray}
which are same as those in the van der Waals liquid-gas system.

\section{Conclusion and discussion}
 Einstein-Horndeski theory is the most general covariant scalar-tensor theory. Stangely enough, it was shown that P-V criticality mimicking the van der Waals liquid-gas phase transition never occurred in the Einstein-Horndeski theory, while such a criticality had been widely found in many gravity theories. In this paper, we have investigated the P-V criticality behaviour in the extended phase space of a black hole system in Einstein-Horndeski theory. Through cautious analysis of the structure of this black hole system in this theory, we have demonstrated that the pressure should be $P=\frac{\zeta}{8\pi \eta}$ rather than the cosmological constant $P=-\frac{1}{8\pi}\Lambda$. Physically, $\frac{\zeta}{\eta}$ uniquely corresponds to the AdS radius, and thus works as an effective cosmological constant. With this new definition of pressure, we have obtained that P-V criticality indeed occurs in the Einstein-Horndeski theory. Furthermore, we have also calculated the critical exponents in this process and checked the scaling laws between these exponents. The exponents follow those in mean field theory, which agrees with previous studies of P-V criticality in many gravity theories. It should be pointed out that $\Lambda_{eff}$ is an effective cosmological constant satisfying $\Lambda_{eff}=-\frac{3}{l_{eff}^2}=-\frac{\zeta}{\eta}$, which is a parameter independent on cosmological constant $\Lambda$. Hence, pressure $P=\frac{\zeta}{8\pi \eta}$ can be totally independent on cosmological constant $\Lambda$, which is a significant difference from previous works. Therefore, our paper is the first to obtain P-V criticality in the two cases with $\Lambda=0$ and $\Lambda>0$, which implicates that the cosmological constant $\Lambda$ may be not a necessary pressure candidate for black holes at the microscopic level.

 For these cases investigated in our paper, a subtlety related to parameters region should be clarified more thoroughly, since a real scalar field $\psi$ with $\psi^2(r)>0$ should exist to make solution (\ref{elepo}) to represent a black hole. More precisely, the corresponding constraint from (\ref{elepo}) is
 \begin{align}
-\frac{4(\zeta+\Lambda \eta)r^4+\eta q^2}{\eta (\zeta r^2+\eta)}=-\frac{4(\frac{\zeta}{\eta}+\Lambda)r^4+q^2}{\zeta r^2+\eta}=-\frac{4(\frac{\zeta}{\eta}+\Lambda)r^4+q^2}{\eta(\frac{\zeta}{\eta} r^2+1)}>0. \label{constrain}
\end{align}
In the case with pressure defined as $P=-\frac{1}{8\pi}\Lambda$, we have chosen $q=1, \Pi^*=\frac{\zeta}{\eta}=0.14$ in Fig.~\ref{fig5}. From this diagram, we can find that $P=-\frac{1}{8\pi}\Lambda<0.005$, and hence $\frac{\zeta}{\eta}+\Lambda>0$. Therefore, the constraint (\ref{constrain}) can be satisfied by setting $\zeta$ and $\eta$ both negative. On the other hand, for the cases with pressure defined as $P=\frac{\zeta}{8\pi \eta}$, we easily obtain $\frac{\zeta}{\eta}+\Lambda>0$ in cases with $\Lambda=0$ and $\Lambda=0.3>0$, while $\frac{\zeta}{\eta}+\Lambda$ is also positive in our case with $\Lambda=-0.3<0$ since $P=\frac{\zeta}{8\pi \eta}>0.05$ in Fig.~\ref{fig33}. Therefore, for these cases with new defined pressure, constraint (\ref{constrain}) can be also satisfied by setting $\zeta$ and $\eta$ both negative.

Note that, most of our calculations are numerical, and whether an analytical investigation is possible on the critical point and exponents is still an open question. On the other hand, there is no uniqueness theorem in Einstein-Horndeski theory, and hence further investigations on critical behaviour and critical exponents of more solutions in Einstein-Horndeski theory will be also an interesting issue. In addition, the critical exponents of P-V criticality obtained in this paper and many cases all follow the ones in mean field theory. Critical exponents beyond mean field theory in some black hole system of gravity theory needs to be further studied, which is related to another underlying question that critical exponents beyond mean field theory depend on special black hole solution or special gravity theory. As we know, some quantity like temperature of black hole just depends on black hole solution, while some quantity like relation between entropy of black hole and area of event horizon depends on gravity theory~\cite{Cai:2008ys}. Finally, another interesting issue is viewed from the scenario of AdS/CFT correspondence. In this scenario, gravitational theory in bulk spacetime corresponds a conformal field theory dwelling on asymptotical anti-de Sitter (AdS) conformal boundary. Since investigations of P-V criticality also have an asymptotic AdS behaviour in bulk spacetime, the corresponding physical process on boundary is also an open issue to be further studied.

\section{Acknowledgements}
Y.P Hu thanks a lot for the discussions with Profs. Hong-Bao Zhang, Hai-Qing Zhang, Hai-Shan Liu and Drs. Run-Qiu Yang, Yihao Yin, and also thanks anonymous referee for constructive and helpful comments. This work is supported by National Natural Science Foundation of China (NSFC) under grant Nos. 11575083, 11565017.

\appendix
\section{First law of thermodynamics of black holes reinvestigated} \label{A}
There are three parameters $\Lambda$, $\zeta$ and $\eta$ in the Lagrangian of action (\ref{action}). Among previous investigations of $P-V$ criticality of asymptotical AdS black hole system, cosmological constant $\Lambda$ is usually considered as a dynamical variable. In our paper, parameter $\zeta$ is also considered as a dynamical variable, while $\eta$ is fixed as a constant for simplicity. Therefore, the first law of thermodynamics of black hole solution in (\ref{FirstLaw}) is further reinvestigated as
\begin{align}
dE=TdS+\Phi_e dQ_e+\Phi_\chi dQ_\chi+V_A d \Lambda +V_B dP_B , \label{NewFirstLaw1}
\end{align}
where energy $E$, temperature $T$, entropy $S$, charge potential $\Phi_e$ , charge $Q_e$, scalar charge $Q_\chi$ and its conjugate potential $\Phi_\chi$ have been explicitly shown in (\ref{Temperature}) and (\ref{Thermo-quanties}). In addition, $V_A$ is the corresponding conjugate to $\Lambda$, while $V_B$ is the corresponding conjugate to $P_B\equiv\zeta/(8\pi\eta)$, and their complicated expressions are
\begin{align}
V_A & =\frac{1}{768 r_h^4}\left(-\frac{24 r_h^5 \zeta  \left(-q^2 \eta +8 r_h^2 \eta +4 r_h^4 (\zeta -\eta
 \Lambda )\right)}{\left(r_h^2 \zeta +\eta \right) (\zeta -\eta  \Lambda )}-\frac{24 r_h^5 \zeta  \left(-q^2 \eta +8 r_h^2
\eta +4 r_h^4 (\zeta -\eta  \Lambda )\right)}{\left(r_h^2 \zeta +\eta \right) (-\zeta +\eta  \Lambda )}\right. \nonumber\\
 & +\frac{1}{\zeta ^3 \eta ^{3/2} \left(r_h^2 \zeta +\eta \right) (\zeta -\eta  \Lambda )^2}\left(-2 r_h \zeta ^{3/2} \eta
 \left(r_h^2 \zeta +\eta \right) \left(2 \sqrt{\zeta } \sqrt{\eta } \left(-48 q^2 r_h^2 \zeta ^2 \eta +q^4 \zeta ^2 \left(-3
r_h^2 \zeta +\eta \right)\right.\right.\right. \nonumber\\
 & \left.\left.+16 r_h^4 \left(r_h^2 \zeta -3 \eta \right) (\zeta -\eta  \Lambda )^2\right)+3 \pi  r_h^3 \left(q^4 \zeta
^4+16 q^2 \zeta ^3 \eta +16 \eta ^2 \left(3 \zeta ^2+2 \zeta  \eta  \Lambda -\eta ^2 \Lambda ^2\right)\right)\right)  \nonumber\\
 & \left.\left.+12 r_h^4 \zeta  \sqrt{\frac{\zeta }{\eta }} \eta ^{3/2} \left(r_h^2 \zeta +\eta \right) \left(q^2 \zeta ^2+12
\zeta  \eta -4 \eta ^2 \Lambda \right) \left(q^2 \zeta ^2+4 \eta  (\zeta +\eta  \Lambda )\right) \text{ArcTan}\left[r_h \sqrt{\frac{\zeta
}{\eta }}\right]\right)\right),
\end{align}
\begin{align}
V_B & =\frac{\pi \eta}{96 r_h^4 \zeta ^3 \left(r_h^2 \zeta +\eta \right)^2} \left(\frac{1}{\zeta -\eta  \Lambda
}6 r_h^3 \zeta ^4 \left(q^2-4 r_h^2+4 r_h^4 \Lambda \right) \left(-q^2 \eta +8 r_h^2 \eta +4 r_h^4 (\zeta -\eta
 \Lambda )\right)\right. \nonumber\\
  & +\frac{1}{-\zeta +\eta  \Lambda }6 r_h^3 \zeta ^4 \left(q^2-4 r_h^2+4 r_h^4 \Lambda \right) \left(-q^2 \eta +8 r_h^2
\eta +4 r_h^4 (\zeta -\eta  \Lambda )\right) \nonumber\\
 & +\frac{1}{\eta ^{3/2} (\zeta -\eta  \Lambda )^2}\left(r_h^2 \zeta +\eta \right) \left(-3 \pi  r_h^4 \sqrt{\zeta } \left(r_h^2
\zeta +\eta \right) \left(q^4 \zeta ^4 (3 \zeta -5 \eta  \Lambda )+8 q^2 \zeta ^2 \eta  \left(\zeta ^2-4 \zeta  \eta  \Lambda -\eta
^2 \Lambda ^2\right)\right.\right. \nonumber\\
 & \left.-16 \eta ^2 \left(\zeta ^3+7 \zeta ^2 \eta  \Lambda +3 \zeta  \eta ^2 \Lambda ^2-3 \eta ^3 \Lambda ^3\right)\right)+2 r_h
\zeta  \sqrt{\eta } \left(q^4 \zeta ^2 \left(2 \eta ^3 \Lambda +3 r_h^4 \zeta ^2 (3 \zeta -5 \eta  \Lambda )+2 r_h^2 \zeta
 \eta  (3 \zeta -5 \eta  \Lambda )\right)\right. \nonumber\\
  & +24 q^2 r_h^2 \zeta ^2 \eta  \left(-4 \eta ^2 \Lambda +r_h^2 \left(\zeta ^2-4 \zeta  \eta  \Lambda -\eta ^2 \Lambda ^2\right)\right)+16
r_h^4 \left(2 r_h^4 \zeta ^3 (\zeta -\eta  \Lambda )^2+2 r_h^2 \zeta  \eta  (\zeta -3 \eta  \Lambda ) (\zeta -\eta
 \Lambda )^2\right. \nonumber\\
  & \left.\left.+3 \eta ^2 \left(\zeta ^3-\zeta ^2 \eta  \Lambda +3 \zeta  \eta ^2 \Lambda ^2-3 \eta ^3 \Lambda ^3\right)\right)\right)+6
r_h^4 \sqrt{\frac{\zeta }{\eta }} \sqrt{\eta } \left(r_h^2 \zeta +\eta \right) \left(q^2 \zeta ^2+4 \eta  (\zeta +\eta
 \Lambda )\right) \left(q^2 \zeta ^2 (3 \zeta -5 \eta  \Lambda )\right. \nonumber\\
  & \left.\left.\left.-4 \eta  \left(\zeta ^2+6 \zeta  \eta  \Lambda -3 \eta ^2 \Lambda ^2\right)\right) \text{ArcTan}\left[r_h \sqrt{\frac{\zeta
}{\eta }}\right]\right)\right). \label{VB}
\end{align}

\section{Proofs of equivalence between investigations on $P-V$ criticality through $P-V$ and $P-r_h$ diagrams} \label{B}
Note that, $P$-$V$ criticality is usually discussed from investigations on pressure function $P(r_h, T)$ instead of function $P(V, T)$ in many previous works~\cite{Kubiznak:2012wp,Gunasekaran:2012dq,Hendi:2012um,Zhao:2013oza,Wei:2012ui,Spallucci:2013osa,Miao:2018fke,Cai:2013qga,Xu:2015rfa,Majhi:2016txt}. In those previous works, thermodynamic volume $V$ usually is $V=\frac{4}{3}\pi r_h^3$, and thus equivalence between investigations on $P$-$V$ criticality through pressure function $P(V, T)$ and $P(r_h, T)$ is easily obtained. In our case, functions $V_A$ and $V_B$ are complicated. Therefore, we will give a simple proof in the following to show equivalence between investigations on $P-V$ criticality through $P-V$ and $P-r_h$ phase diagrams in our cases, i.e. obtaining the same critical point and critical exponents during using $P-V$ or $P-r_h$ phase diagrams. For simplicity and consistence with constraint from real scalar filed in (\ref{elepo}), we fix constant parameter $\eta$ as $\eta=-1$.
\subsection{Same critical point}
Besides the three parameters $\Lambda$, $\zeta$ and $\eta$ in action, there are in fact two other parameters $M$ and $q$ from black hole solution (\ref{elepo}). For simplicity, $\eta$ has been fixed as a constant $\eta=-1$, therefore, now the whole parameters are just $\Lambda$, $\zeta$, $M$ and $q$. In this subsection, we will give a simple proof to show that the critical point are same by using pressure function $P(V, T)$ or $P(r_h,T)$. Without loss of generality, we take pressure function $P(V_B,T)$ with new defined pressure case into account. In this case, $\zeta$ and $M$ are two dynamical variables, while $q$ and $\Lambda$ are fixed parameters.

The critical point from pressure function $P(V_B, T)$ is obtained from
\begin{eqnarray}
\left(\frac{\partial P}{\partial V_B}\right)_T & =&0,~~\left(\frac{\partial ^2P}{\partial V_B{}^2}\right)_T =0, \label{Eqcritical}
\end{eqnarray}
while the critical point from pressure function $P(r_h, T)$ is
\begin{eqnarray}
\left(\frac{\partial P}{\partial r_h}\right)_T=0,~~\left(\frac{\partial ^2P}{\partial r_h{}^2}\right)_T=0.\label{NewEqcritical}
\end{eqnarray}
In the following, we will prove that (\ref{Eqcritical}) and (\ref{NewEqcritical}) obtain the same critical point. Here a key point is using $V_B$ and $T$ as the two dynamical variables, and hence $P\left(r_h,T\right) =P\left(r_h(V_B,T),T\right)=P(V_B, T)$, thus we can obtain
\begin{eqnarray}
\left(\frac{\partial P}{\partial V_B}\right)_T&=& \left(\frac{\partial P}{\partial r_h}\right)_T\left(\frac{\partial r_h}{\partial V_B}\right)_T,\\
\left(\frac{\partial ^2P}{\partial V_B{}^2}\right){}_T&=&\left[\left(\frac{\partial ^2P}{\partial r_h{}^2}\right){}_T \left(\frac{\partial r_h}{\partial V_B}\right)_T+\left(\frac{\partial P}{\partial r_h}\right)_T\left(\frac{\partial }{\partial r_h}\right)_T\left(\frac{\partial r_h}{\partial V_B}\right)_T\right]\left(\frac{\partial r_h}{\partial V_B}\right)_T.
\end{eqnarray}
We will prove that $\left(\frac{\partial r_h}{\partial V_B}\right)_T$ is nonzero around the critical point in the following subsection. Therefore, (\ref{Eqcritical}) and (\ref{NewEqcritical}) are easily found to be equivalent, and hence they obtain the same critical point.

\subsection{Same critical exponents}
In our paper, P-V criticality just exists in the case with new defined pressure. Therefore, we just focus on this case in the following discussions. For the critical exponent $\alpha$, it is deduced from
\begin{align}
C_{V_B} & =\left(\frac{\text{TdS}}{\text{dT}}\right)_{V_B},
\end{align}
or
\begin{align}
C_v & =\left(\frac{\text{TdS}}{\text{dT}}\right)_v,
\end{align}
while $v$ is specific volume $v=2 r_h$, and entropy $S$ has been obtained in (\ref{Entropy}) as a function $S(P, r_h)$. In our cases, we can prove that the isopyknic heat capacity $C_{V_B}$ or $C_v$ at the critical point are both constant. Therefore, we have $\alpha=0$.

For critical exponents $\beta$, $\gamma$ and $\delta$, we need expand pressure function $P(V_B,T)$ around the critical point like $P(r_h,T)$ in (\ref{Pexpansion}).
In the case with pressure function $P(V_B,T)$, expansion around the critical point is
\begin{eqnarray}
P_B(V_B, T)&=&P_{\text{Bc}}+\left[\left(\frac{\partial P_B}{\partial T}\right)_{V_B}\right]_c\left(T-T_c\right)+\frac{1}{2}\left[\left(\frac{\partial ^2P_B}{\partial
T^2}\right)_{V_B}\right]_c\left(T-T_c\right){}^2+\left[\frac{\partial ^2P_B}{\partial T\partial V_B}\right]_c\left(V_B-V_{\text{Bc}}\right)\left(T-T_c\right)\nonumber\\
&+&\frac{1}{6}\left[\left(\frac{\partial
^3P_B}{\partial V_B{}^3}\right)_T\right]_c\left(V_B-V_{\text{Bc}}\right){}^3+...,
\end{eqnarray}
where $B$ has been also indexed in pressure P to correspond $V_B$ to make discussions more clearly hereafter, i.e. $P_B(V_B,T)=P(V_B,T)$. Note that, for the coefficients, we can obtain
\begin{eqnarray}
\left(\frac{\partial P_B}{\partial T}\right)_{V_B}=\left(\frac{\partial P_B}{\partial r_h}\right)_T\left(\frac{\partial r_h}{\partial T}\right)_{V_B}+\left(\frac{\partial P_B}{\partial T}\right)_{r_h}
\end{eqnarray}
where $P_B(r_h,T)=P_B(r_h(V_B,T),T)=P_B(V_B,T)$ has been used, while other coefficients are as following
\begin{align}
\left(\frac{\partial }{\partial T}\right)_{V_B}\left(\frac{\partial P_B}{\partial V_B}\right)_T & =\left(\frac{\partial }{\partial T}\right)_{V_B}\left[\left(\frac{\partial
P_B}{\partial r_h}\right)_T\left(\frac{\partial r_h}{\partial V_B}\right)_T\right] \nonumber \\
& =\left(\frac{\partial r_h}{\partial V_B}\right)_T\left(\frac{\partial }{\partial T}\right)_{V_B}\left(\frac{\partial P_B}{\partial r_h}\right)_T+\left(\frac{\partial
P_B}{\partial r_h}\right)_T\left(\frac{\partial }{\partial T}\right)_{V_B}\left(\frac{\partial r_h}{\partial V_B}\right)_T \nonumber\\
& =\left[\left(\frac{\partial ^2P_B}{\partial r_h{}^2}\right){}_T \left(\frac{\partial r_h}{\partial T}\right)_{V_B}+\frac{\partial ^2P_B}{\partial
T \partial r_h}\right]\left(\frac{\partial r_h}{\partial V_B}\right)_T+\left(\frac{\partial P_B}{\partial r_h}\right)_T\left(\frac{\partial }{\partial
T}\right)_{V_B}\left(\frac{\partial r_h}{\partial V_B}\right)_T,
\end{align}
and
\begin{eqnarray}
\left(\frac{\partial ^2P_B}{\partial T^2}\right){}_{V_B} &=&\left(\frac{\partial }{\partial T}\right)_{V_B}\left[\left(\frac{\partial
P_B}{\partial r_h}\right)_T\left(\frac{\partial r_h}{\partial T}\right)_{V_B}+\left(\frac{\partial P_B}{\partial T}\right)_{r_h}\right] \nonumber\\
& =&\left(\frac{\partial r_h}{\partial T}\right)_{V_B}\left(\frac{\partial }{\partial T}\right)_{V_B}\left(\frac{\partial P_B}{\partial r_h}\right)_T+\left(\frac{\partial
P_B}{\partial r_h}\right)_T\left(\frac{\partial }{\partial T}\right)_{V_B}\left(\frac{\partial r_h}{\partial V_B}\right)_T+\left(\frac{\partial }{\partial
T}\right)_{V_B}\left(\frac{\partial P_B}{\partial T}\right)_{r_h} \nonumber\\
& =&\left[\left(\frac{\partial ^2P_B}{\partial r_h{}^2}\right){}_T\left(\frac{\partial r_h}{\partial T}\right)_{V_B}+\frac{\partial ^2P_B}{\partial
T \partial r_h}\right]\left(\frac{\partial r_h}{\partial T}\right)_{V_B}+\left(\frac{\partial P_B}{\partial r_h}\right)_T\left(\frac{\partial }{\partial
T}\right)_{V_B}\left(\frac{\partial r_h}{\partial V_B}\right)_T+\left(\frac{\partial }{\partial T}\right)_{V_B}\left(\frac{\partial P_B}{\partial
T}\right)_{r_h},\nonumber\\
\left(\frac{\partial ^3P_B}{\partial V_B{}^3}\right){}_T & =&\left(\frac{\partial r_h}{\partial V_B}\right)_T\left(\frac{\partial }{\partial
r_h}\right)_T\left(\frac{\partial ^2P_B}{\partial V_B{}^2}\right){}_T \nonumber\\
& =&\left[\left(\frac{\partial }{\partial r_h}\right)_T\left(\left[\left(\frac{\partial ^2P_B}{\partial r_h{}^2}\right){}_T\left(\frac{\partial
r_h}{\partial V_B}\right)_T+\left(\frac{\partial P_B}{\partial r_h}\right)_T\left(\frac{\partial }{\partial r_h}\right)_T\left(\frac{\partial r_h}{\partial
V_B}\right)_T\right]\left(\frac{\partial r_h}{\partial V_B}\right)_T\right)\right]\left(\frac{\partial r_h}{\partial V_B}\right)_T.
\end{eqnarray}

At the critical point, we can find that these coefficients are nonzero. For example, we obtain these coefficients in the case with $q=0.3, \Lambda=-0.3$ as
\begin{eqnarray}
\left[\left(\frac{\partial P_B}{\partial T}\right)_{V_B}\right]_c &=&\left[\left(\frac{\partial P_B}{\partial T}\right)_{r_h}\right]_c=1.8617,~~
\left[\left(\frac{\partial ^2P_B}{\partial T^2}\right)_{V_B}\right]_c = 2 \left[\frac{\partial^2P_B}{\partial T \partial r_h}\left(\frac{\partial r_h}{\partial T}\right)_{V_B}\right]_c+\left[\left(\frac{\partial ^2P_B}{\partial T^2}\right)_{r_h}\right]_c=-9.700,\nonumber\\
\left[\frac{\partial^2P_B}{\partial T\partial V_B}\right]_c&=&\left[\frac{\partial^2P_B}{\partial T \partial r_h}\left(\frac{\partial r_h}{\partial V_B} \right)_T\right]_c=-17.1345,~~
\left[\left(\frac{\partial ^3P_B}{\partial V_B{}^3}\right)_T\right]_c=\left[\left(\frac{\partial ^3P_B}{\partial r_h{}^3}\right)_T\left(\frac{\partial r_h}{\partial V_B}\right)_T^3\right]_c=-4093.1, \label{NewCoefficents}
\end{eqnarray}
while several numerical values at the critical point have been used
\begin{eqnarray}
&&\left[\left(\frac{\partial V_B}{\partial r_h}\right)_{P_B}\right]_c \approx 0.3964,~
\left[\left(\frac{\partial V_B}{\partial P_B}\right)_{r_h}\right]_c \approx -0.1581,~
\left[\left(\frac{\partial T}{\partial r_h}\right)_{P_B}\right]_c \approx 5.9369\times 10^{-15},~
\left[\left(\frac{\partial T}{\partial P_B}\right)_{r_h}\right]_c \approx 0.5372,\nonumber\\
&&\left[\left(\frac{\partial r_h}{\partial V_B}\right)_T\right]_c \approx 2.5230,~
\left[\left(\frac{\partial P_B}{\partial V_B}\right)_T\right]_c \approx -2.7886\times 10^{-14},~
\left[\left(\frac{\partial r_h}{\partial T}\right)_{V_B}\right]_c \approx 0.7424,~
\left[\left(\frac{\partial P_B}{\partial T}\right)_{V_B}\right]_c \approx 1.8617,\nonumber\\
&&\left[\left(\frac{\partial P_B}{\partial T}\right)_{r_h}\right]_c \approx 1.8617,~
\left[\left(\frac{\partial ^2P_B}{\partial T\partial r_h}\right)\right]{}_c \approx -6.7913,~
\left[\left(\frac{\partial ^3P_B}{\partial r_h{}^3}\right){}_T\right]{}_c \approx -254.853,~
\left[\left(\frac{\partial ^2P_B}{\partial T^2}\right){}_{r_h}\right]{}_c \approx 0.3839.\nonumber
\end{eqnarray}
Since these coefficients in (\ref{NewCoefficents}) are non-zero, therefore, similar to discussions in the text and Refs. \cite{Kubiznak:2012wp,Xu:2015rfa,Majhi:2016txt}, the three critical exponents $\beta$, $\gamma$ and $\delta$ are same.

Note that, quantities in the above expressions are not obviously obtained. Therefore, in the following, we will give a detailed calculation for $\left(\frac{\partial r_h}{\partial V_B}\right)_T$ as an example. The starting point is that we have obtained temperature and thermodynamic volume as functions of $r_h$ and $P_B$, i.e. $T\left(r_h,P_B\right)$ and $V_B\left(r_h,P_B\right)$ in (\ref{Temperature}) and (\ref{VB}). Since there are just two dynamical variables in our cases, in principle we can also use $T$ and $V_B$ as the two new dynamical variables. Therefore, from $T\left(r_h,P_B\right)\equiv T\left(r_h(T,V_B),P_B(T,V_B)\right)$ and $V_B\left(r_h,P_B\right)\equiv V_B\left(r_h(T,V_B),P_B(T,V_B)\right)$, we deduce the following key relations
\begin{align}
\text{dT} & =\left(\frac{\partial T}{\partial r_h}\right)_{P_B}\text{dr}_h+\left(\frac{\partial T}{\partial P_B}\right)_{r_h}\text{dP}_B \nonumber\\
& =\left(\frac{\partial T}{\partial r_h}\right)_{P_B}\left[\left(\frac{\partial r_h}{\partial V_B}\right)_T\text{dV}_B+\left(\frac{\partial r_h}{\partial
T}\right)_{V_B}\text{dT}\right]+\left(\frac{\partial T}{\partial P_B}\right)_{r_h}\left[\left(\frac{\partial P_B}{\partial V_B}\right)_T\text{dV}_B+\left(\frac{\partial
P_B}{\partial T}\right)_{V_B}\text{dT}\right] \nonumber\\
& =\left[\left(\frac{\partial T}{\partial r_h}\right)_{P_B}\left(\frac{\partial r_h}{\partial V_B}\right)_T+\left(\frac{\partial T}{\partial P_B}\right)_{r_h}\left(\frac{\partial
P_B}{\partial V_B}\right)_T\right]\text{dV}_B+\left[\left(\frac{\partial T}{\partial r_h}\right)_{P_B}\left(\frac{\partial r_h}{\partial T}\right)_{V_B}+\left(\frac{\partial
T}{\partial P_B}\right)_{r_h}\left(\frac{\partial P_B}{\partial T}\right)_{V_B}\right]\text{dT},
\end{align}
\begin{align}
\text{dV}_B &=\left(\frac{\partial V_B}{\partial r_h}\right)_{P_B}\text{dr}_h+\left(\frac{\partial V_B}{\partial P_B}\right)_{r_h}\text{dP}_B \nonumber\\
& =\left(\frac{\partial V_B}{\partial r_h}\right)_{P_B}\left[\left(\frac{\partial r_h}{\partial V_B}\right)_T\text{dV}_B+\left(\frac{\partial
r_h}{\partial T}\right)_{V_B}\text{dT}\right]+\left(\frac{\partial V_B}{\partial P_B}\right)_{r_h}\left[\left(\frac{\partial P_B}{\partial V_B}\right)_T\text{dV}_B+\left(\frac{\partial
P_B}{\partial T}\right)_{V_B}\text{dT}\right] \nonumber\\
& =\left[\left(\frac{\partial V_B}{\partial r_h}\right)_{P_B}\left(\frac{\partial r_h}{\partial v_B}\right)_T+\left(\frac{\partial V_B}{\partial
P_B}\right)_{r_h}\left(\frac{\partial P_B}{\partial V_B}\right)_T\right]\text{dV}_B+\left[\left(\frac{\partial V_B}{\partial r_h}\right)_{P_B}\left(\frac{\partial
r_h}{\partial T}\right)_{V_B}+\left(\frac{\partial V_B}{\partial P_B}\right)_{r_h}\left(\frac{\partial P_B}{\partial T}\right)_{V_B}\right]\text{dT}.
\end{align}
From which, four equations should be satisfied
\begin{align}
\left\{
             \begin{array}{lr}
             \left(\frac{\partial V_B}{\partial r_h}\right)_{P_B}\left(\frac{\partial r_h}{\partial V_B}\right)_T+\left(\frac{\partial V_B}{\partial
P_B}\right)_{r_h}\left(\frac{\partial P_B}{\partial V_B}\right)_T=1, &  \\
             \left(\frac{\partial T}{\partial r_h}\right)_{P_B}\left(\frac{\partial r_h}{\partial V_B}\right)_T+\left(\frac{\partial T}{\partial P_B}\right)_{r_h}\left(\frac{\partial
P_B}{\partial V_B}\right)_T=0. &
             \end{array}
\right.
\end{align}

\begin{align}
\left\{
             \begin{array}{lr}
             \left(\frac{\partial V_B}{\partial r_h}\right)_{P_B}\left(\frac{\partial r_h}{\partial T}\right)_{V_B}+\left(\frac{\partial V_B}{\partial P_B}\right)_{r_h}\left(\frac{\partial
P_B}{\partial T}\right)_{V_B}=0, &  \\
             \left(\frac{\partial T}{\partial r_h}\right)_{P_B}\left(\frac{\partial r_h}{\partial T}\right)_{V_B}+\left(\frac{\partial T}{\partial P_B}\right)_{r_h}\left(\frac{\partial
P_B}{\partial T}\right)_{V_B}=1. &
             \end{array}
\right.
\end{align}
therefore, we can finally obtain
\begin{align}
\left\{
             \begin{array}{lr}
             \left(\frac{\partial r_h}{\partial V_B}\right)_T=\frac{1}{\left(\frac{\partial V_B}{\partial r_h}\right)_{P_B}-\left(\frac{\partial V_B}{\partial
P_B}\right)_{r_h}\frac{\left(\frac{\partial T}{\partial r_h}\right)_{P_B}}{\left(\frac{\partial T}{\partial P_B}\right)_{r_h}}}, &  \\
             \left(\frac{\partial P_B}{\partial V_B}\right)_T=\frac{1}{\left(\frac{\partial V_B}{\partial P_B}\right)_{r_h}-\left(\frac{\partial V_B}{\partial
r_h}\right)_{P_B}\frac{\left(\frac{\partial T}{\partial P_B}\right)_{r_h}}{\left(\frac{\partial T}{\partial r_h}\right)_{P_B}}}, &  \\
            \left(\frac{\partial r_h}{\partial T}\right)_{V_B}=\frac{1}{\left(\frac{\partial T}{\partial r_h}\right)_{P_B}-\left(\frac{\partial T}{\partial
P_B}\right)_{r_h}\frac{\left(\frac{\partial V_B}{\partial r_h}\right)_{P_B}}{\left(\frac{\partial V_B}{\partial P_B}\right)_{r_h}}}, & \\
            \left(\frac{\partial P_B}{\partial T}\right)_{V_B}=\frac{1}{\left(\frac{\partial T}{\partial P_B}\right)_{r_h}-\left(\frac{\partial T}{\partial
r_h}\right)_{P_B}\frac{\left(\frac{\partial V_B}{\partial P_B}\right)_{r_h}}{\left(\frac{\partial v_B}{\partial r_h}\right)_{P_B}}}. & \\
             \end{array}
\right.
\end{align}
Therefore, from the known functions $T\left(r_h,P_B\right)$ and $V_B\left(r_h,P_B\right)$ in (\ref{Temperature}) and (\ref{VB}), we can easily obtain $\left(\frac{\partial r_h}{\partial V_B}\right)_T$ and other quantities around the critical point.


\begin{thebibliography}{99}


\bibitem{Bardeen}
J. M. Bardeen, B. Carter and S. W. Hawking, Commun. Math. Phys. 31, 161 (1973); J.D. Bekenstein, Physical Review D7, 2333, (1973).
\bibitem{Hawking}
S. W. Hawking, Commun. Math. Phys. 43, 199 (1975).
\bibitem{HP}
S. W. Hawking and D. N. Page, Commun. Math. Phys. 87, 577 (1983).

\bibitem{Gibbons:1976pt}
  G.~W.~Gibbons and M.~J.~Perry,
  Proc.\ Roy.\ Soc.\ Lond.\ A {\bf 358}, 467 (1978).

\bibitem{Hu:2017nzw}
  Y.~P.~Hu, F.~Pan and X.~M.~Wu,
  Phys.\ Lett.\ B {\bf 772}, 553 (2017)
  [arXiv:1703.08599 [gr-qc]].



\bibitem{witten}
E.~Witten,
  Adv.\ Theor.\ Math.\ Phys.\  {\bf 2}, 505 (1998)
  [hep-th/9803131].


\bibitem{Chamblin:1999hg}
  A.~Chamblin, R.~Emparan, C.~V.~Johnson and R.~C.~Myers,
  Phys.\ Rev.\ D {\bf 60}, 104026 (1999)
  [hep-th/9904197].

\bibitem{Kubiznak:2012wp}
  D.~Kubiznak and R.~B.~Mann,
  JHEP {\bf 1207}, 033 (2012)
  [arXiv:1205.0559 [hep-th]].

\bibitem{Gunasekaran:2012dq}
  S.~Gunasekaran, R.~B.~Mann and D.~Kubiznak,
  JHEP {\bf 1211}, 110 (2012)
  [arXiv:1208.6251 [hep-th]].
  N.~Altamirano, D.~Kubiznak and R.~B.~Mann,
  Phys.\ Rev.\ D {\bf 88}, no. 10, 101502 (2013)
  [arXiv:1306.5756 [hep-th]].


\bibitem{Hendi:2012um}
  S.~H.~Hendi and M.~H.~Vahidinia,
  Phys.\ Rev.\ D {\bf 88}, no. 8, 084045 (2013)
  [arXiv:1212.6128 [hep-th]];
  S.~H.~Hendi and A.~Dehghani,
  Eur.\ Phys.\ J.\ C {\bf 79}, no. 3, 227 (2019)
  [arXiv:1811.01018 [gr-qc]].

\bibitem{Zhao:2013oza}
  R.~Zhao, H.~H.~Zhao, M.~S.~Ma and L.~C.~Zhang,
  Eur.\ Phys.\ J.\ C {\bf 73}, 2645 (2013)
  [arXiv:1305.3725 [gr-qc]].

\bibitem{Wei:2012ui}
  S.~W.~Wei and Y.~X.~Liu,
  Phys.\ Rev.\ D {\bf 87}, no. 4, 044014 (2013)
  [arXiv:1209.1707 [gr-qc]];
  S.~W.~Wei and Y.~X.~Liu,
  Phys.\ Rev.\ Lett.\  {\bf 115}, no. 11, 111302 (2015)
  Erratum: [Phys.\ Rev.\ Lett.\  {\bf 116}, no. 16, 169903 (2016)]
  [arXiv:1502.00386 [gr-qc]].

\bibitem{Spallucci:2013osa}
  E.~Spallucci and A.~Smailagic,
  Phys.\ Lett.\ B {\bf 723}, 436 (2013)
  [arXiv:1305.3379 [hep-th]].


\bibitem{Miao:2018fke}
  Y.~G.~Miao and Z.~M.~Xu,
  Phys.\ Rev.\ D {\bf 98}, no. 8, 084051 (2018)
  [arXiv:1806.10393 [hep-th]].


\bibitem{Cai:2013qga}
  R.~G.~Cai, L.~M.~Cao, L.~Li and R.~Q.~Yang,
  JHEP {\bf 1309}, 005 (2013)
  [arXiv:1306.6233 [gr-qc]].



\bibitem{Xu:2015rfa}
  J.~Xu, L.~M.~Cao and Y.~P.~Hu,
  Phys.\ Rev.\ D {\bf 91}, no. 12, 124033 (2015)
  [arXiv:1506.03578 [gr-qc]].



\bibitem{Majhi:2016txt}
  B.~R.~Majhi and S.~Samanta,
  Phys.\ Lett.\ B {\bf 773}, 203 (2017)
  [arXiv:1609.06224 [gr-qc]];
  K.~Bhattacharya and B.~R.~Majhi,
  Phys.\ Rev.\ D {\bf 95}, no. 10, 104024 (2017)
  [arXiv:1702.07174 [gr-qc]];
  K.~Bhattacharya, B.~R.~Majhi and S.~Samanta,
  Phys.\ Rev.\ D {\bf 96}, no. 8, 084037 (2017)
  [arXiv:1709.02650 [gr-qc]].



\bibitem{Kastor:2009wy}
  D.~Kastor, S.~Ray and J.~Traschen,
  Class.\ Quant.\ Grav.\  {\bf 26}, 195011 (2009)
  [arXiv:0904.2765 [hep-th]].


  \bibitem{vacexp}
  G. W. Gibbons, R. Kallosh and B. Kol, Phys. Rev. Lett. 77, 4992 (1996) [hepth/
9607108]; J. D. E. Creighton and R. B. Mann, Phys. Rev. D 52, 4569 (1995) [gr-qc/9505007].




 \bibitem{horn}
 G. W. Horndeski, Int. J. Theor. Phys. 10, 363 (1974).

\bibitem{Deffayet:2011gz}
  C.~Deffayet, X.~Gao, D.~A.~Steer and G.~Zahariade,
  Phys.\ Rev.\ D {\bf 84}, 064039 (2011)
  [arXiv:1103.3260 [hep-th]].

\bibitem{Gao:2014soa}
  X.~Gao,
  Phys.\ Rev.\ D {\bf 90}, 081501 (2014)
  [arXiv:1406.0822 [gr-qc]];
  X.~Gao,
  Phys.\ Rev.\ D {\bf 90}, 104033 (2014)
  [arXiv:1409.6708 [gr-qc]];
  X.~Gao and Z.~b.~Yao,
  arXiv:1806.02811 [gr-qc].




\bibitem{bhhorn}
  A.~Anabalon, A.~Cisterna and J.~Oliva,
  Phys.\ Rev.\ D {\bf 89}, 084050 (2014)
  [arXiv:1312.3597 [gr-qc]];
  M.~Minamitsuji,
  Phys.\ Rev.\ D {\bf 89}, 064017 (2014)
  [arXiv:1312.3759 [gr-qc]].



\bibitem{Cisterna:2014nua}
  A.~Cisterna and C.~Erices,
  Phys.\ Rev.\ D {\bf 89}, 084038 (2014)
  [arXiv:1401.4479 [gr-qc]].

\bibitem{Feng:2015wvb}
  X.~H.~Feng, H.~S.~Liu, H.~L\"u and C.~N.~Pope,
  Phys.\ Rev.\ D {\bf 93}, no. 4, 044030 (2016)
  [arXiv:1512.02659 [hep-th]].

\bibitem{Miao:2016aol}
  Y.~G.~Miao and Z.~M.~Xu,
  Eur.\ Phys.\ J.\ C {\bf 76}, no. 11, 638 (2016)
  [arXiv:1607.06629 [hep-th]].


\bibitem{Cai:2008ys}
  R.~G.~Cai, L.~M.~Cao and Y.~P.~Hu,
  JHEP {\bf 0808}, 090 (2008)
  [arXiv:0807.1232 [hep-th]].


\end{thebibliography}
\end{document}